%% file: main.tex
\documentclass[conference]{IEEEtran}
\IEEEoverridecommandlockouts
\usepackage{cite}
\usepackage{amsmath,amssymb,amsfonts}
\usepackage{textcomp}
\usepackage{xcolor}
\usepackage[a4paper, total={184mm,239mm}]{geometry}
\input{packages}

\input{macro}

\def\BibTeX{{\rm B\kern-.05em{\sc i\kern-.025em b}\kern-.08em
    T\kern-.1667em\lower.7ex\hbox{E}\kern-.125emX}}
\graphicspath{{./figs/}}
\usepackage{geometry}
\begin{document}

\title{SimPhony: A Heterogeneous Electronic-\underline{Pho}tonic AI System Cross-La\underline{y}er \underline{Sim}ulation Framework with Accurate, Versatile Device, Circuit, Layout Representation and Modeling}

\title{\SIM: A Device-Circuit-Architecture Cross-Layer Modeling and \underline{Sim}ulation Framework for Heterogeneous Electronic-\underline{Pho}to\underline{n}ic AI S\underline{y}stem
\vspace{-10pt}
}
\author
{
Ziang Yin$^1$, 
Meng Zhang$^2$, 
Amir Begovic$^2$, 
Rena Huang$^2$, 
Jeff Zhang$^1$,
Jiaqi Gu$^1$\\
$^1$Arizona State University, $^2$Rensselaer Polytechnic Institute\\
{\small \emph{jiaqigu@asu.edu}}
\vspace{-5pt}
}

\maketitle
\input{doc/0abstract}
\input{doc/1intro}

\input{doc/2prelim}

\input{doc/3method}

\input{doc/4results}
\input{doc/5conclusion}

\input{main.bbl}

\end{document}

%% file: packages.tex
\usepackage[utf8]{inputenc} %
\usepackage[T1]{fontenc}    %
\usepackage{pifont}
\usepackage[colorlinks,
    linkcolor=red,citecolor=citecolor,urlcolor=blue]{hyperref}       %

\usepackage{graphicx}
\usepackage{todonotes}
\usepackage{amssymb}
\usepackage{threeparttable}
\usepackage{multirow}
\usepackage{amsmath}
\usepackage{bm}
\usepackage{float}
\usepackage{amsthm}
\usepackage{soul}
\usepackage[noend]{algpseudocode}
\usepackage{enumitem} %
\usepackage[subrefformat=parens,labelformat=parens]{subfig}

\usepackage{xspace}

%% file: macro.tex
\definecolor{citecolor}{RGB}{34,139,34}
\definecolor{mydarkblue}{rgb}{0,0.08,1}
\definecolor{mydarkgreen}{rgb}{0.02,0.6,0.02}
\definecolor{mydarkred}{rgb}{0.8,0.02,0.02}
\definecolor{mydarkorange}{rgb}{0.40,0.2,0.02}
\definecolor{mypurple}{RGB}{111,0,255}
\definecolor{myred}{rgb}{1.0,0.0,0.0}
\definecolor{mygold}{rgb}{0.75,0.6,0.12}
\definecolor{myblue}{rgb}{0,0.2,0.8}
\definecolor{mydarkgray}{rgb}{0.,0.2,0.2}

\definecolor{lightred}{RGB}{255,235,235}
\definecolor{lightgreen}{RGB}{235,255,235}
\definecolor{lightblue}{RGB}{235,235,255}
\definecolor{lightcyan}{RGB}{235,255,255}
\definecolor{lightmagenta}{RGB}{255,235,255}
\definecolor{lightyellow}{RGB}{255,255,235}

\definecolor{qxkcolor}{RGB}{215,235,255}
\definecolor{softmaxcolor}{RGB}{230,235,255}
\definecolor{probxvcolor}{RGB}{255,255,235}

\definecolor{topkcolor}{RGB}{255,235,235}
\definecolor{zecolor}{RGB}{255,255,235}
\definecolor{dynacolor}{RGB}{235,255,255}

\definecolor{reviewcolor}{RGB}{0,0,200}

\theoremstyle{plain}

\theoremstyle{definition}

\algdef{SE}[DOWHILE]{Do}{doWhile}{\algorithmicdo}[1]{\algorithmicwhile\ #1}%

\newcommand{\squishlist}{
 \begin{list}{$\bullet$}
  { \setlength{\itemsep}{0pt}
     \setlength{\parsep}{3pt}
     \setlength{\topsep}{3pt}
     \setlength{\partopsep}{0pt}
     \setlength{\leftmargin}{1.5em}
     \setlength{\labelwidth}{1em}
     \setlength{\labelsep}{0.5em} } }
     
\newcommand{\squishend}{
  \end{list}  }

\newcommand{\SIM}{\texttt{SimPhony}\xspace}
\newcommand{\SIMDev}{\texttt{SimPhony-DevLib}\xspace}
\newcommand{\SIMArch}{\texttt{SimPhony-Arch}\xspace}
\newcommand{\SIMSim}{\texttt{SimPhony-Sim}\xspace}
\newcommand{\LT}{\texttt{LT}\xspace}

%% file: doc/0abstract.tex
\begin{abstract}
\label{abstract}
Electronic-photonic integrated circuits (EPICs) offer transformative potential for next-generation high-performance AI but require interdisciplinary advances across devices, circuits, architecture, and design automation. 
The complexity of hybrid systems makes it challenging even for domain experts to understand distinct behaviors and interactions across design stack.
The lack of a flexible, accurate, fast, and easy-to-use EPIC AI system simulation framework significantly limits the exploration of hardware innovations and system evaluations on common benchmarks. 
To address this gap, we propose \SIM\footnote{We open-source our codes at \href{https://github.com/ScopeX-ASU/SimPhony}{https://github.com/ScopeX-ASU/SimPhony}}, a
cross-layer modeling and simulation framework for heterogeneous electronic-photonic AI systems. 
\SIM offers a platform that enables 
(1) generic, extensible hardware topology representation that supports heterogeneous multi-core architectures with diverse photonic tensor core designs;
(2) optics-specific dataflow modeling with unique multi-dimensional parallelism and reuse beyond spatial/temporal dimensions;
(3) data-aware energy modeling with realistic device responses, layout-aware area estimation, link budget analysis, and bandwidth-adaptive memory modeling; and
(4) seamless integration with model training framework for hardware/software co-simulation.
By providing a unified, versatile, and high-fidelity simulation platform, \SIM enables researchers to innovate and evaluate EPIC AI hardware across multiple domains, facilitating the next leap in emerging AI hardware.

\end{abstract}

%% file: doc/1intro.tex
\section{Introduction}
\label{sec:Introduction}
Heterogeneous electronic-photonic integrated circuits (EPICs) are emerging as a next-generation platform for high-performance artificial intelligence (AI) computing.
Demonstrated optical neural networks (ONNs) showcase breakthroughs in their performance and efficiency~\cite{NP_NATURE2017_Shen, NP_Nature2021_Feldmann, NP_ACS2022_Feng, NP_HPCA2024_Zhu, NP_ISCA2021_Shiflett, NP_NaturePhotonics2021_Shastri, NP_Science2024_Xu,NP_JLT2024_Ning,NP_JSTQE2020_Totovic}.
Many research focuses on materials, devices, and circuits to overcome technological barriers in photonic AI.
Some cross-layer co-design~\cite{NP_TCAD2020_Gu, NP_DAC2020_Gu, NP_DATE2020_Gu,NP_ISVLSI2022_Banerjee,NP_ICCAD24_Gu} and architecture optimization~\cite{NP_ISCA2021_Shiflett, NP_HPCA2024_Zhu, NP_DATE2019_Liu, NP_DATE2020_Zokaee, NP_JAP2024_Zhang,NP_HPCA2020_Shiflett,NP_ISCA2024_Demirkiran} research have scaled these systems for real-world AI tasks.
Exploring the full stack of EPIC AI systems demands expertise across physics, device design, analog circuits, system architecture, electronic-photonic design automation (EPDA), and AI algorithms, as system-level evaluation is critical to understanding innovations at each individual design layer.

However, the complexity of such hybrid system makes it challenging even for experts to grasp the behavior of each component and its interactions across software and hardware.
Key challenges include:
\ding{202}~\textbf{Lack of Unified Representation of Distinct PTC Circuit Topology}:~
Photonic tensor cores (PTCs) employ diverse optical principles for matrix computation, e.g., magnitude/phase modulation, wavelength/mode/time-division multiplexing (WDM, MDM, TDM), interference, diffraction, etc.
Such abundant design flexibility creates various PTC circuit topologies, e.g., weight bank~\cite{NP_SciRep2017_Tait}, triangular mesh~\cite{NP_PHYSICAL1994_Reck, NP_NATURE2017_Shen}, rectangular mesh~\cite{NP_Optica2018_Clements}, butterfly mesh~\cite{NP_ACS2022_Feng}, crossbar~\cite{NP_Nature2021_Feldmann, NP_HPCA2024_Zhu, NP_JAP2024_Zhang}, single WDM link~\cite{NP_Nature2021_Xu}, etc.
Prior simulators are modified from digital tools~\cite{NP_arXiv2024_Andrulis} that only support array-like computing architectures and cannot represent diverse PTC topologies.
\ding{203}~\textbf{Lack of Support for Optics-Specific Dataflow and Parallelism}:
EPIC AI systems involve parallelism and resource sharing beyond temporal/spatial dimensions.
The combination of optical broadcasting, hierarchical accumulation, and multi-dimensional reuse patterns results in a complex dataflow.
The additional optical dimensions, like magnitude/phase, wavelength, polarization, and modes, further complicate the design space.
Thus, a flexible framework tailored to optics-specific needs is required.
\ding{204}~\textbf{Lack of Accurate Model-Circuit-Layout Co-Modeling}:
Unlike digital AI accelerators, analog systems integrate models tightly with devices/circuits, leading to inaccuracies in energy modeling due to unawareness of real workload data and precise device settings.
Beyond simple analytical models, there is a strong need to incorporate rigorous simulations and even chip measurements into energy analysis.
Additionally, existing EPIC design tools overlook the layout when estimating the chip area. 
Previous methods~\cite{NP_ASPDAC2020_Gu, NP_HPCA2024_Zhu,NP_APLML2024_Gu} aggregate device footprints, leading to an underestimate of area. 
Therefore, a fast yet accurate layout estimator is essential for more reliable area analysis.

In this work, we present \SIM, an open-source cross-layer modeling and simulation framework for heterogeneous EPIC AI systems. 
Built with a customized EPIC device library, \SIM enables the hierarchical construction of heterogeneous photonic architectures from arbitrary PTC circuit topologies. 
Integrated with the ONN training library, \SIM enables end-to-end simulation, including workload extraction, memory construction, and dataflow generation, while accurately analyzing system latency, data-aware energy, and layout-aware area.

Our main contributions are as follows,
\begin{itemize}
    \item We introduce \SIM, a cross-layer simulation framework for EPIC AI systems, enabling end-to-end accurate performance and efficiency analysis.
    \item \textbf{Unified PTC Representation}: 
    Utilizing our customized device library, we design a hierarchical netlist-based circuit representation that generates arbitrary PTC topologies with automatically derived scaling rules and critical paths. 
    \item \textbf{Photonics-Specific Dataflow Handling}: \SIM accommodates multi-dimensional parallelism and hierarchical accumulation, effectively integrating the unique characteristics and dataflow of photonic computing hardware.
    \item \textbf{Accurate System Modeling}: \SIM provides accurate energy analysis based on real workload data and realistic device power models, along with accurate area analysis featuring auto-generated layout estimation.
\end{itemize}

%% file: doc/2prelim.tex
\vspace{-5pt}
\section{Preliminary}
\label{sec:prelim}
\subsection{Photonic Tensor Core Taxonomy and Modeling Challenges}
\label{sec:PTC}
The diverse properties of PTC designs result in variations in circuit topology, devices, and operational principles, affecting speed and power.
Table~\ref{tab:EPICDifference} categorizes PTCs by expressivity, computing mechanism, operand range, and reconfiguration speed~\cite{NP_JLT2024_Ning}.
For \underline{expressivity}, universal PTCs can perform arbitrary matrix multiplication, e.g., micro-ring (MRR) arrays~\cite{NP_SciRep2017_Tait}, while subspace PTCs support only a subset of static linear transformation~\cite{NP_ACS2022_Feng}.
Based on the \underline{numerical range of input operands}, full-range PTCs compute arbitrary values in one shot, whereas subspace coherent PTCs (e.g., Butterfly meshes) require two differential computations for full-range input/output.
Incoherent PTCs, e.g., MRR arrays~\cite{NP_SciRep2017_Tait}, also require two computations to handle full-range inputs, while non-volatile phase change material (PCM) crossbar arrays~\cite{NP_Nature2021_Feldmann} need up to four cycles.
\underline{Reconfiguration speed} is another key factor to determine its supported dataflow and operations. 
Thermo-optic MZI arrays require matrix decomposition and thermal tuning to reconfigure the weights, leading to a delay of $\mu s$ to $ms$ and limiting them to weight-stationary dataflows, unsuitable for dynamic self-attention.
Dynamic PTCs~\cite{NP_HPCA2024_Zhu, NP_JAP2024_Zhang} use high-speed modulators for real-time matrix switching, enabling dynamic tensor products and output-static dataflows.
Developing a modeling framework that accurately models complex system performance is challenging.
\input{tables/EPIC_designs}

PTCs also feature varied topologies beyond conventional crossbars, such as triangular~\cite{NP_PHYSICAL1994_Reck} and butterfly meshes~\cite{NP_ACS2022_Feng}.
Architectural features like optical broadcast, multi-dimensional sharing~\cite{NP_HPCA2024_Zhu}, analog-domain accumulation, and electrical-optical (E-O) interfaces further complicate system description and accurate modeling.

\vspace{-3pt}
\subsection{Photonic Accelerator Simulation Tools}
Prior photonic AI hardware is evaluated based on internal closed-source analytical device modeling and behavior-level modeling~\cite{NP_ACS2022_Feng,NP_DAC2022_Gu,NP_ICCAD2021_Li, NP_DATE2019_Liu, NP_DATE2020_Zokaee}, unaware of the numerical values in actual workloads.
CimLoop is a recent analog NN simulator; its photonic version~\cite{NP_arXiv2024_Andrulis} has showcased one architecture Albireo~\cite{NP_ISCA2021_Shiflett}.
However, it %
focuses on architectural parameters (e.g., memory and dataflow), and lacks support for flexible PTC construction with photonics-specific dataflow and parallelism. 
Its oversimplified device modeling and the complicated interface of the Timeloop-based framework make it challenging for device and circuit researchers to explore hardware optimization efficiently.
Furthermore, previous efforts have not interfaced with ONN training tools at the application level, lacking an integrated ONN training and \emph{device/circuit-centric} system modeling infrastructure, which is needed to model EPIC AI system efficiency and performance.

%% file: tables/EPIC_designs.tex
\begin{table}
\caption{\small PTC taxonomy~\cite{NP_JLT2024_Ning}: PTC designs with distinct properties in operand range and reconfiguration frequency (\emph{Reconfig}) with different \# of forward required to obtain full-range output.}
\resizebox{\columnwidth}{!}{
\begin{tabular}{c|cc|cc|c|c}
\hline
\multirow{2}{*}{\textbf{EPIC Designs}} & \multicolumn{2}{c|}{\textbf{Operand A}}                             & \multicolumn{2}{c|}{\textbf{Operand B}}                             & \multirow{2}{*}{\textbf{\begin{tabular}[c]{@{}c@{}}Forward \\ Method\end{tabular}}} & \multirow{2}{*}{\textbf{\#Forwards}} \\ \cline{2-5}
                                       & \textbf{Range}                                 & \textbf{Reconfig} & \textbf{Range}                                 & \textbf{Reconfig} &                                                                                     &                                                                                                     \\ \hline
MZI Array~\cite{NP_NATURE2017_Shen}                              & $\mathbb{R}$                   & Dynamic            & $\mathbb{R}$                   & Static             & Direct                                                                              & 1                                                                                                \\ \hline
Butterfly Mesh~\cite{NP_TCAD2020_Gu}                   & $\mathbb{R}$                   & Dynamic            & $\mathbb{C}$                   & Static             & Pos-Neg                                                                             & 1                                                                                                \\ \hline
MRR Array~\cite{NP_SciRep2017_Tait}                        & $\mathbb{R^+}$ & Dynamic            & $\mathbb{R}$                   & Dynamic            & Direct                                                                              & 2                                                                                               \\ \hline
PCM crossbar~\cite{NP_APR2020_Miscuglio}                    & $\mathbb{R^+}$ & Dynamic            & $\mathbb{R^+}$ & Static             & Direct                                                                              & 4                                                                                        \\ \hline
TeMPO~\cite{NP_JAP2024_Zhang}                                  & $\mathbb{R}$                   & Dynamic            & $\mathbb{R}$                   & Dynamic            & Direct                                                                              & 1                                                                                                \\ \hline
\end{tabular}
}
\label{tab:EPICDifference}
\vspace{-8pt}
\end{table}

%% file: doc/3method.tex
\vspace{-3pt}
\section{\SIM: EPIC AI System Modeling and Simulation Framework}
\label{sec:method}
\input{fig_tex/fig_overview}
\vspace{-3pt}
Figure~\ref{fig:FrameworkOverview} summarizes the simulation flow and key components of our proposed \SIM framework.
It contains a customized electronic-photonic device library \SIMDev and supports flexible, hierarchical, and parametric modeling of EPIC AI system architecture \SIMArch.
Integrated with \texttt{TorchONN} model training toolkit, our simulation system \SIMSim handles photonics-specific dataflow with bandwidth-adaptive memory hierarchy, data-aware power estimation, link budget analysis, and layout-aware chip area analysis.

\vspace{-13pt}
\subsection{\SIMDev: Comprehensive and Customizable Electronic-Photonic Device Library}
\label{sec:DevLib}
One key novelty of our simulation framework is the support of a comprehensive and customizable device library \SIMDev, which lays the foundation for flexible architecture construction and accurate system modeling.

\noindent\textbf{Modeling Granularity}: \SIMDev device modeling is based on experimental data reported, ensuring an accurate representation of device characteristics.
Specifically, the photonic device power models are obtained through Lumerical HEAT simulation or experimental measurements that are aware of actual device configuration, ensuring fast and accurate cost modeling.

\noindent\textbf{Modeling Approach}: We organize key components into electrical and optical categories, including a variety of high-performance photonic and electronic devices.
Comprehensive device information is provided in detail to support accurate simulations of area, power, latency, and link budget. 
Devices from foundry PDKs can be plugged in. 
Our device library supports flexible scaling with different technology nodes, port numbers, working conditions, and more.
For example, %
digital-to-analog converters (DACs) in our library support power scaling with customized sampling rates and bit resolutions, enabling power optimization via gating or quantization. 
On photonic device front, e.g., the electro-optic Mach-Zehnder modulator (MZM) is widely used for high-speed encoding. 
To achieve precise performance modeling, we collect various properties such as spatial size, bandwidth, insertion loss, modulation efficiency, static power, extinction ratio, testing bitwidth, and more.

\input{fig_tex/fig_node_def}
\input{fig_tex/fig_parametric_tempo}

\subsection{\SIMArch: Hierarchical, Parametric Heterogeneous EPIC AI System Architecture Builder}
\label{sec:SIMArch}
To enable flexible construction of heterogeneous multi-core architectures with diverse PTC designs and dataflows, we introduce \SIMArch, a hierarchical, parametric architecture builder.

Existing simulators focus on dataflow-centric architectural modeling with fixed PTC designs, often neglecting device/circuit details. 
This limits their suitability for fundamental device/circuit-level customization, underscoring the need for a universal representation to unify diverse PTC designs.

To enable flexible PTC construction, we customize a \emph{netlist} representation in \SIMArch to describe devices as instances and port connectivity as directed 2-pin nets.
Unlike electrical circuit netlists with undirected multi-pin nets, PTCs require directed 2-pin nets to capture the directional optical signal flow.

Key observations of PTC design patterns inspire us to use \emph{modular circuit construction}, avoiding manual scripting of large netlists.
This allows us to define \emph{a minimal building block} denoted as \emph{node}, e.g., a dot-product unit shown in Fig.~\ref{fig:NodeDef}, and \emph{build the circuit according to specific scaling rules} with a user-defined node connection topology.
A weighted directed acyclic graph (DAG) is generated based on the node topology, as shown in Fig.~\ref{fig:NodeDAG}.
The topology and insertion loss-based edge weights are essential in link budget analysis and layout-aware area estimation.
This universal, hierarchical netlist interface also enables potential SPICE simulation and physical design as a future extension.

In the following, we provide two case studies of how to construct representative PTC architectures in a parametric style.

\noindent\textbf{Case Study 1: Dynamic Array-style Tensor Cores TeMPO~\cite{NP_JAP2024_Zhang,NP_HPCA2024_Zhu}}.~
The first case study demonstrates how to construct an array-style PTC architecture, TeMPO~\cite{NP_JAP2024_Zhang,NP_HPCA2024_Zhu}, designed for dynamic time-multiplexed tensor products, as shown in Fig.~\ref{fig:ParametricTeMPO}.
We first define the \emph{architecture parameters}: $R$ tiles, each containing $C$ cores, with $H \times W$ dot-product nodes per core performing parallel computations.
Then, we define the structure of the minimum building block, i.e., the dot-product unit, denoted as a \emph{node}.
A \emph{node netlist} is used to describe the 6-device circuit topology using \emph{directed} 2-pin nets to represent the waveguide connections and signal flow, shown in Fig.~\ref{fig:NodeDef}.
To efficiently span the multi-core architecture without manually detailing every connection, we define \emph{scaling rules} applied to each node and describe inter-node connections.
This approach supports \textbf{parametric generation} of the architecture, enabling \textbf{automatic analysis} of area, power, and link budget.
For example, the device area/power estimation engine will trace the netlist to count the number of devices considering \emph{hardware sharing}. 
There are $RCHW$ total nodes for parallel dot-product.
As the output of $C$ cores in a tile are \emph{in-situ} accumulated, integrators/ADCs can be shared and thus scaled by $CHW$.
MZM group A encodes one matrix operand and can be broadcast to $R$ tiles. 
Thus, the input encoders, i.e., DAC A and MZM A, are scaled by $RH$.
These scaling rules are expressed as \emph{customizable symbolic expressions} in circuit description files, enabling user-defined reuse styles to suit specific designs.
The link budget analyzer will trace the netlist and auto-extract the longest path from the weighted DAG of the hierarchical netlist. 
Edge weights are automatically assigned to the insertion loss of incident vertices to match the parametric architecture settings, e.g., edge pointing to $i4$ stores $(CW-1)\times$ the loss of device $i4$.

\noindent\textbf{Case Study 2: Static Mesh-style MZI Array~\cite{NP_NATURE2017_Shen, NP_Optica2018_Clements}}.~
Besides array style, \SIMArch can handle \emph{more challenging mesh-style} PTCs, e.g., Clements-style MZI meshes in Fig.~\ref{fig:ParametricMZIArray}.
The matrix is encoded by singular value decomposition ($U\Sigma V)$, where unitary matrices $U/V$ are decomposed to interleaved MZI meshes.
To parametrically generate MZI meshes, we follow the \emph{node} description to define a single-MZI node-$U$, node-$\Sigma$, and node-$V$.
Our flexible \emph{scaling rule} allows users to independently build unitary matrices by scaling node-$U/V$ by $RCH(H-1)/2$ times and build diagonal by scaling node-$\Sigma$ by $RC \min(H,W)$ times, which is \emph{not representable by prior simulators that are based on arrays}. 
Details of the critical path and hardware sharing are omitted here, as they follow similar principles to Case Study 1.

\vspace{-3pt}
\subsection{\SIMSim: EPIC AI System Simulation Flow}
\label{sec:Simulation}
\SIMSim is an end-to-end simulation flow, including NN model conversion and workload extraction to memory simulation and analysis of latency, area, and energy cost.

\subsubsection{ONN Model Conversion and Workload Extraction}
\label{sec:Workload}
For digital DNN accelerator simulation, model training and hardware mapping are sufficiently decoupled.
In contrast, analog mixed-signal EPIC AI systems require cross-layer co-design, resulting in closely coupled model training, conversion, mapping, and architecture simulation processes.
Our \SIMSim system can seamlessly interface with an open-source ONN training library, \texttt{TorchONN}~\cite{NP_NeurIPS2021_Gu}, that supports various customized ONN types.
A digital DNN will be first \textbf{converted to its analog optical version with layer-wise conversion}, e.g., \texttt{Conv2d} to \texttt{TeMPOConv2d}.
The converted model will be trained with device non-ideality, quantization, pruning, and various co-design methodologies to ensure accuracy, robustness, and energy efficiency.
The converted ONN model will be parsed by \SIMSim to extract the detailed workload configuration for each layer, including input/weight size, input/weight/output bitwidth, pruning mask, scaling factors, \textbf{actual weight values}, etc.
Weight values can have different modes, e.g., matrix values, normalized device transmissions, phase shifts, or even control voltages, which are useful for \textbf{precise value-aware power modeling}.
Convolution, linear, and attention layers will be converted to general matrix multiplication (GEMM) representations.
Other less computation-intensive layers are offloaded to electrical processors and omitted here for simplicity.
With a layer-to-arch mapping configuration, we enable the flexibility to \textbf{map different layers to different types of sub-architectures} based on their compatibility and efficiency considerations, enabling heterogeneous computing paradigms.

\input{fig_tex/photonic_sharing}

\subsubsection{Photonics-Specific Dataflow and Latency Analysis}
\label{sec:PSD}
Besides the support for standard dataflow for GEMM, e.g., weight/input/output stationary, here, we emphasize unique photonic-specific mapping and parallelism in \SIMSim.

\noindent\textbf{Multi-Dimension Parallelism and Hierarchical Accumulation}.~
Beyond the spatial and temporal dimensions of electrical hardware, optical systems have more physical dimensions for parallel computing and hardware sharing, e.g., spectral, polarization, modes, etc.
Figure~\ref{fig:PhotonicParallelism} illustrates a partitioned GEMM workload mapped to a multi-core TeMPO architecture with multi-dimensional parallelism and hierarchical accumulation.
As shown in the nested loop representation, multiple wavelengths are used for spectral parallel summation, photocurrents from $C$ cores are aggregated for analog-domain parallel summation, spatial parallelism is used for parallel outer product, temporal integration is used for analog sequential summation, and the partial sum is further sequentially accumulated digitally in the local buffer.
Based on the mapping and parallelism, we will \emph{derive the system execution latency in units of cycles}.

\noindent\textbf{Latency Penalty for Range-Restricted PTCs}.~As highlighted in Section~\ref{sec:prelim} and Table~\ref{tab:EPICDifference}, PTCs are constrained in their \emph{numerical range representation due to device modulation features}, leading to \emph{varying processing times to complete full-range computations}.
We denote the number of iterations to acquire full-range output as $I$.
Especially for PTCs that can only encode positive inputs/weights, four times the execution cost is required to realize full-range output, i.e., $I$=4.
\SIM will automatically analyze the tensor core property based on input/weight/output encoding properties and generate the corresponding dataflow with $I\times$ latency penalty. 

\noindent\textbf{PTC Reconfiguration Latency Penalty}.~
In weight-stationary PTCs, loading new weights is sometimes bottlenecked by slow device reprogramming rather than memory loading. 
For instance, thermo-optic (TO) devices have a thermal time constant of ~10 µs, and writing to phase change material (PCM) cells incur a delay of over 100 ns. 
\SIMSim automatically analyzes reprogramming latency and \textbf{applies corresponding cycle penalty whenever weight loading causes circuit reconfiguration delays exceeding one clock cycle}, e.g., 500 cycles per switch for 100 ns reconfiguration delay at 5 GHz.
The total latency of mapping one layer is $\tau_{tot}=\tau_{\text{load-input/weight}}+\tau_{\text{write-out}}+I(\tau_{\text{comp}}+\tau_{\text{reconfig}})$.

\subsubsection{Bandwidth-Adaptive Memory Hierarchy Modeling}
\label{sec:Memory}
One of the key points to enabling photonics advantage in AI computing is \emph{sufficient data movement bandwidth and latency-hiding} techniques.
Note that we focus on memory bandwidth analysis and assume on-chip and cross-chiplet interconnects provide sufficient data transaction bandwidth, especially when optical interconnects are available for multi-Tbps signal fanout/broadcast bandwidth~\cite{NN_SIGCOMM21_KHANI}.
\SIMArch adopts a four-level memory hierarchy consisting of off-chip High Bandwidth Memory (HBM), Global Buffer (GLB), Local Buffer (LB), and Register File (RF). 
Each memory level stores operands A, B, and the output in progressively smaller sizes:i.e., the entire model at the HBM level, a single layer at the GLB level, the processing matrix dimensions at the LB level, and data for a single cycle at the RF level.
The bandwidth of LB ($BW_{LB}$) and RF ($BW_{RF}$) must accommodate the architecture's single-cycle processing throughput, calculated as $BW_{LB}, BW_{RF} \geq  Bytes Per Cycle \times f$,
where $f$ is the PTC operating clock frequency.
GLB's bandwidth ($BW_{GLB}$) is calculated as $BW_{GLB} = Max Layer Size \cdot f/(N_p \cdot D_P \cdot M_p)$,
where the partitioned matrix dimensions are $N_{p}$, $D_{p}$, and $M_{p}$.
We first profile the \textbf{maximum bandwidth requirement ($\widehat{BW}$) for all sub-architectures} based on the GLB demand per cycle, considering data sharing and broadcast in a specific dataflow.
To enable full utilization of the computing cores \emph{without memory bottleneck}, we adopt a SoTA \textbf{multi-block SRAM design to meet the bandwidth demand} ---
A multi-block GLB can boost its bandwidth proportional to the number of blocks~\cite{NP_HPCA2024_Zhu, NP_JAP2024_Zhang}.
\SIMSim automatically searchs for the minimum number of required GLB blocks, $\#\text{ of Blocks} = \tau_{GLB}\cdot \widehat{BW}/(b_{\text{bus}} \cdot 8)$,
where $\tau_{GLB}$ is the fastest cycle simulated by CACTI~\cite{CACTI7}, and $b_{bus}$ is the buswidth in bits.

\subsubsection{Link Budget Analysis}
\label{sec:LinkBudget}
Link budget analysis is important for the photonic systems to profile the critical-path insertion loss and derive the laser source power requirement and optical signal-to-noise ratio (SNR).
To obtain the IL on the critical path, we leverage our constructed hierarchical, weighted DAG representation of the architecture in Fig.~\ref{fig:ParametricTeMPO},~\ref{fig:ParametricMZIArray}.
From a given source node, i.e., laser, to a destination node, i.e., photodetector (PD), we will use the graph's longest path to get the critical path IL. 
Given the PD sensitivity $S$, to differentiate $b_{in}$-bit input levels with a target bit error rate, we can derive the lowest laser power required~\cite{NP_HPCA2024_Zhu, NP_ACS2022_Feng},
\vspace{-5pt}
\begin{equation}
\label{eq:LaserPower}
   \small
    P_{laser} = \frac{10^{(S + IL) / 10} \cdot 2^{b_{in}}}{\eta_{WPE}} \cdot \frac{1}{1 - 0.1^{ER / 10}},
    \vspace{-5pt}
\end{equation}
where $\eta_{WPE}$ is the laser wall plug efficiency. 
Non-ideal modulation extinction ratio $ER$ is also considered a power penalty to recover the full modulation range~\cite{NP_SciRep2017_Tait}.
\input{fig_tex/fig_data_dependent_energy}

\subsubsection{Data-Dependent Device-Response-Aware Energy Analysis}
To accurately model the energy of EPIC AI systems, \SIMSim captures data access energy via CACTI-simulated memory energy and computing energy based on \noindent\emph{actual operand values and realistic device power modeling}.

\noindent\underline{For data access cost}, we derive the memory access size ($D_{mem}$) for off-chip HBM, GLB, LB, and RF based on the dataflow analysis and accumulate the energy cost with CACTI-simulated access energy per bit $e_{mem}$, i.e., $E_{mem}=\sum_{mem\sim\{HBM,GLB,LB,RF\}}e_{mem}D_{mem}$.

\noindent\underline{For computing cost}, \SIMSim supports \textbf{data-dependent} energy analysis with \textbf{accurate device power modeling}, shown in Fig.~\ref{fig:DDEnergy}.
For analog hardware, the encoded data determines the device configuration, significantly impacting its power.
Hence, \SIM \textbf{accumulates the energy over cycles based on the values of the real operands}.
This approach enables accurate energy profiling with \emph{fine-grained power gating from ONN pruning}~\cite{NP_ICCAD24_Gu}.
Device power modeling is also critical.
Default library power references, like $P_{\pi}$ for phase shifters (PSs), often overestimate actual power, while analytical models can be overly ideal.
As shown in Fig.~\ref{fig:DDEnergy}, 
\SIMSim supports \textbf{customized power models} using analytical, simulation, or chip testing data for power estimation with different fidelity.

\subsubsection{Layout-Aware Chip Area Analysis}
\label{sec:Area}
\input{fig_tex/fig_layout_area}
The chip area is crucial for fabrication and packaging costs and guiding design optimization.
\SIMSim supports fast, realistic layout-aware area estimation. Unlike previous methods that simply sum all device footprints, 
\SIMSim either takes in a user-defined bounding box or \textbf{automatically generates a signal-flow-aware floorplan}, as shown in Fig.~\ref{fig:LayoutArea}. 
\SIM sets the placement site width to fit the longest device and attempts to \emph{hide} other devices beneath it. 
The \textbf{floorplan follows the device's topological order} from the netlist to adhere to the minimum bending rule in PIC placement, accounting for user-defined device/node spacing.
This approach closely matches the real layout area and can be potentially extended to interface with PIC placement tools.

%% file: fig_tex/fig_overview.tex
\begin{figure}
    \centering
    \vspace{-3pt}
    \includegraphics[width=\columnwidth]{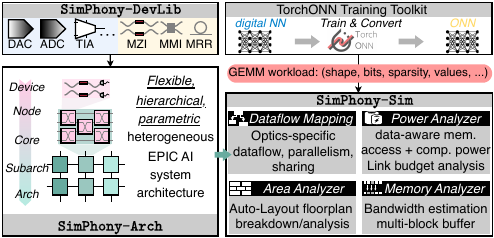}
    \vspace{-10pt}
    \caption{\small Overview of our proposed \SIM framework.}
    \label{fig:FrameworkOverview}
    \vspace{-10pt}
\end{figure}

%% file: fig_tex/fig_node_def.tex
\begin{figure}
    \centering
    \vspace{-10pt}
    \subfloat[]{\includegraphics[width=0.5\columnwidth]{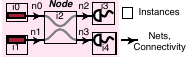}
    \label{fig:NodeDef}} 
    \hspace{15pt}
    \subfloat[]{\includegraphics[width=0.26\columnwidth]{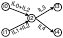}
    \label{fig:NodeDAG}}
    \vspace{-5pt}
    \caption{\small (a) Node-level circuit topology definition and (b) corresponding weighted direct acyclic graph (DAG) representation.}
    \label{fig:NodeTopo}
    \vspace{-10pt}
\end{figure}

%% file: fig_tex/fig_parametric_tempo.tex
\begin{figure}
    \centering
    \subfloat[]{\includegraphics[width=\columnwidth]{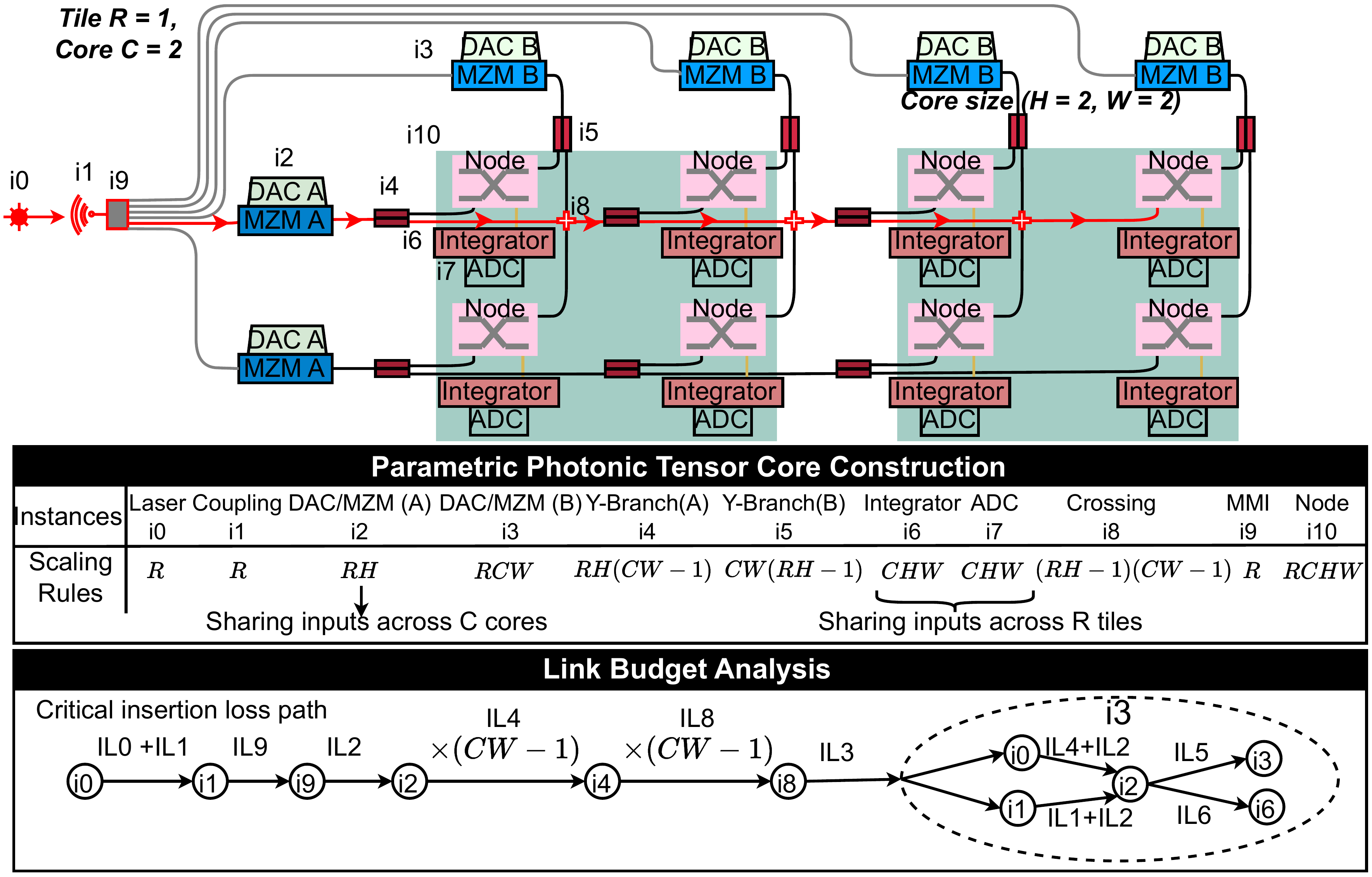}
    \label{fig:ParametricTeMPO}}\\
    \vspace{-10pt}
    \subfloat[]{\includegraphics[width=\columnwidth]{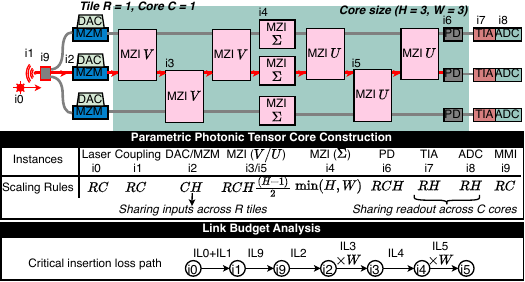}
    \label{fig:ParametricMZIArray}}
    \vspace{-5pt}
    \caption{\small Case studies of constructing parametric multi-core accelerator architecture (a) dynamic array-style TeMPO~\cite{NP_JAP2024_Zhang} and (b) static mesh-style MZI arrays~\cite{NP_NATURE2017_Shen}.
    The circuits can be generated by scaling up the smallest node with a user-defined scaling rule.
    The critical path with the highest insertion loss can be auto-derived.}
    \vspace{-12pt}
    \label{fig:ParametricDesign}
\end{figure}

%% file: fig_tex/photonic_sharing.tex
\begin{figure}[H]
    \centering
    \vspace{-10pt}
    \includegraphics[width=0.95\columnwidth]{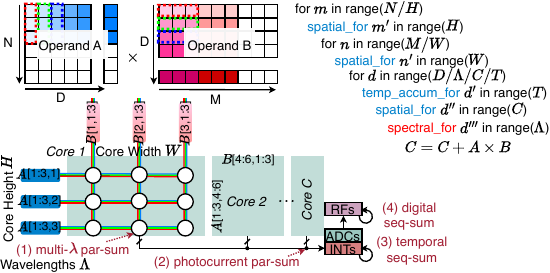}
    \caption{\small Mapping blocking GEMM to output-stationary PTC with spatial/spectral parallelism and hierarchical accumulation.}
    \vspace{-10pt}
    \label{fig:PhotonicParallelism}
\end{figure}

%% file: fig_tex/fig_data_dependent_energy.tex
\begin{figure}
    \centering
    \includegraphics[width=0.81\columnwidth]{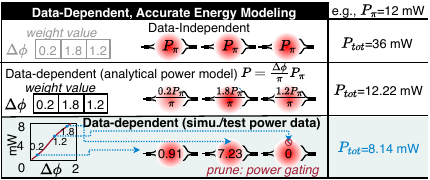}
    \vspace{-3pt}
    \caption{\small \SIMSim supports accurate, data-dependent energy modeling with simulated/measured device power data.}
    \label{fig:DDEnergy}
    \vspace{-13pt}
\end{figure}

%% file: fig_tex/fig_layout_area.tex
\begin{figure}
    \centering
    \includegraphics[width=0.85\columnwidth]{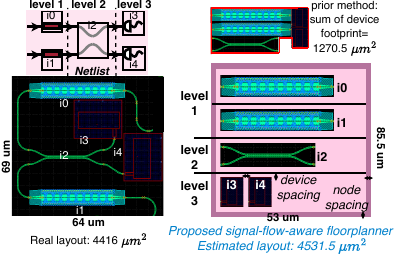}
    \vspace{-5pt}
    \caption{\small Proposed signal flow-aware row-based floorplan for accurate layout-aware area estimation.
    }
    \label{fig:LayoutArea}
    \vspace{-10pt}
\end{figure}

%% file: doc/4results.tex
\vspace{-5pt}
\section{Evaluation Results}
\label{sec:result}
\input{fig_tex/fig_sim_valid}
\input{fig_tex/fig_sim_valid_dota}
\subsection{\SIM Validation}
\label{sec:valid}
\vspace{-3pt}
we validate our simulation results by comparing them with architectural evaluation results reported in previous work.
It is important to clarify that, in the absence of system-level experimental demonstrations of multi-core photonic AI chips, we compare our results with prior architecture simulation results.
We re-emphasize that individual \emph{component data used are backed by experimental measurements}, and area estimates are based on real chip layouts. 
The functionality and accuracy of the ONN model are ensured by the TorchONN training framework, which is beyond the scope of our architecture performance modeling tool.

\noindent\textbf{GEMM Workloads}: To validate the simulation results of \SIM for GEMM, we compare the simulated area and energy metrics on a (280$\times$28)$\times$(28$\times$280) GEMM task against the reference values reported in the original TeMPO paper~\cite{NP_JAP2024_Zhang} in Fig.~\ref{fig:Valid} with the following architecture settings. 
We set the core width and height to 4 and the number of tiles and cores per tile to 2. 
The area~\ref{fig:AreaValid} and energy~\ref{fig:EnergyValid} breakdown from \SIM match the reference results from TeMPO paper.

\noindent\textbf{Dynamic Transformer Workloads}: For Transformers with self-attention operations, we validate \SIM's area and energy results against Lightening-Transformer~\cite{NP_HPCA2024_Zhu} (\LT), shown in Fig.~\ref{fig:DOTAValid}. 
we simulate BERT-Base~\cite{NN_BERT2018_Devlin} with a single (224$\times$224) ImageNet-1K~\cite{NN_imagenet2009} image. 
We adopt \LT's settings: 4 tiles, 2 cores per tile, each core sized 12$\times$12, with 12 wavelengths operating at 5 GHz. 
Since \LT provides only power breakdowns, we report power instead of energy. 
SimPhony accurately reproduces the chip area when appropriate scaling factors are applied, matching \LT's reported values. 
Minor deviations in photonic core and memory areas stem from differences in core area calculations, memory simulations, and device spacing. Our layout-aware estimator effectively generates realistic core areas. Device power aligns with \LT's breakdown except for memory, where deviations result from different technology nodes used in memory simulation (\SIM uses CACTI-45 nm~\cite{CACTI7}; \LT uses PCACTI-14 nm~\cite{PCACTI}) and SRAM port/bus differences.

\subsection{\SIM Use Cases}
To show the capability of \SIM, we study multiple design examples by sweeping PTC architectural parameters in \SIM to gain design insights and demonstrate their impacts on system performance and efficiency.
All simulations discussed below have a 4$\times$4 core size with 2 tiles and 2 cores per tile.

\subsubsection{Multi-Wavelength Parallelism}
As discussed in Section~\ref{sec:PSD}, optical systems allow parallelism beyond spatial and temporal dimensions. 
However, data encoding in these dimensions requires extra power. 
Figure~\ref{fig:Wavelength} examines TeMPO~\cite{NP_JAP2024_Zhang} under varying wavelength settings on a (280$\times$28)$\times$(28$\times$280) GEMM task while scaling MZM and laser sources with the number of wavelengths.
Increased wavelengths enhance parallelism, speeding up computation and reducing energy for components that do not scale with wavelength.
However, the MZM's energy remains constant as the number of MZMs scale with \# of wavelengths.
\input{fig_tex/fig_bit_precision}

\subsubsection{Bitwidth Representation vs. Energy Consumption}
To investigate how the ADC/DAC bit precision impacts the system power, in Fig.~\ref{fig:BitPrecision}, we sweep the energy for different tensor bitwidth. 
The results show a clear trend of increasing energy with higher bits. 
Users can leverage bitwidth simulation results to find the sweet spot for optimal efficiency.
\input{fig_tex/fig_data_aware}

\subsubsection{Layout-Aware and Data-Dependent Modeling}
The exploration of layout-aware area estimation and data-dependent modeling is shown in Fig.~\ref{fig:PhysicalAware}.
The layout-unaware method underestimates the node area by 72\%, while our floorplan estimation enables accurate area estimation compared to the real layout.
In data-awareness evaluation, we focus on a weight-static PTC \texttt{SCATTER}~\cite{NP_ICCAD24_Gu} where weight values impact the phase shifter's power. 
By counting PS power based on real weight values, the PS energy decreases from 0.0537 $\mu$J to 0.0215 $\mu$J with an approximate power model.
If a rigorous device power model is used, the energy is further reduced to 0.0209 $\mu$J with a substantial 60\% reduction.

\input{fig_tex/fig_vgg8_mapping}
\subsubsection{Heterogeneous Mapping}
Lastly, we show \SIM's capability for heterogeneous architecture modeling with layer-to-sub-architecture mapping in Fig.~\ref{fig:Vgg8}. 
The convolutional layers of VGG-8 are mapped to \texttt{SCATTER}~\cite{NP_ICCAD24_Gu}, while the linear layers are mapped to MZI meshes~\cite{NP_NATURE2017_Shen}, while two sub-architectures share the same on-chip memory hierarchy. 
This hybrid architecture modeling showcases \SIM's flexibility in integrating hybrid systems and enabling fine-grained workload-to-hardware mapping. 
As a future extension, \SIM can be extended to enable automated design space exploration that combines the strengths of different photonic computing architectures in heterogeneous systems for diverse AI workloads.

%% file: fig_tex/fig_sim_valid.tex
\begin{figure}
    \centering
    \vspace{-10pt}
    \subfloat[]{\includegraphics[width=0.49\columnwidth]{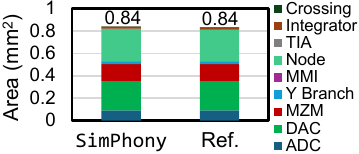}
    \label{fig:AreaValid}} 
    \subfloat[]{\includegraphics[width=0.455\columnwidth]{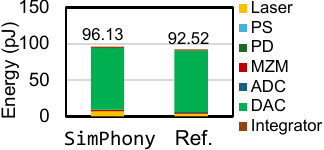}
    \label{fig:EnergyValid}}
    \vspace{-5pt}
    \caption{\small \SIM results validated on a (280$\times$28)$\times$(28$\times$280) GEMM task w/ TeMPO~\cite{NP_JAP2024_Zhang}. (a) area (b) energy breakdown.}
    \label{fig:Valid}
    \vspace{-13pt}
\end{figure}

%% file: fig_tex/fig_sim_valid_dota.tex
\begin{figure}[t]
    \centering
    \vspace{-10pt}
    \subfloat[]{\includegraphics[width=0.51\columnwidth]{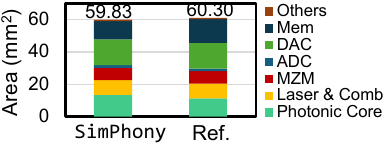}
    \label{fig:DOTAAreaValid}
    } 
    \subfloat[]{\includegraphics[width=0.43\columnwidth]{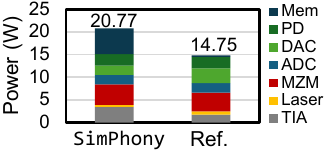}
    \label{fig:DOTAPowerValid}
    }
    \vspace{-5pt}
    \caption{\small \SIM results validated on BERT-Base(ImageNet) w/ Lightening-Transformer~\cite{NP_HPCA2024_Zhu}. (a) area (b) power breakdown.}
    \label{fig:DOTAValid}
    \vspace{-15pt}
\end{figure}

%% file: fig_tex/fig_bit_precision.tex
\begin{figure}
    \centering
    \vspace{-10pt}
    \subfloat[]
    {\includegraphics[width=0.4\columnwidth]{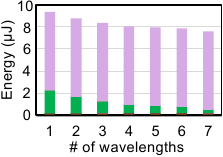}
    \label{fig:Wavelength}}
    \hspace{5pt}
    \subfloat[]
    {\includegraphics[width=0.525\columnwidth]{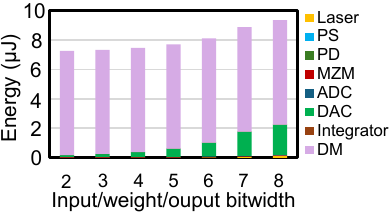}
    \label{fig:BitPrecision}}
    \vspace{-5pt}
    \caption{\small Sweep (a) \# of wavelengths (b) bitwidth on TeMPO arch~\cite{NP_JAP2024_Zhang} and (280$\times$28)$\times$(28$\times$280) GEMM. \emph{DM}: data movement.}
    \label{fig:ParamSweep}
    \vspace{-10pt}
\end{figure}

%% file: fig_tex/fig_data_aware.tex
\begin{figure}
    \centering
    \vspace{-10pt}
    \subfloat[]{\includegraphics[width=0.52\columnwidth]{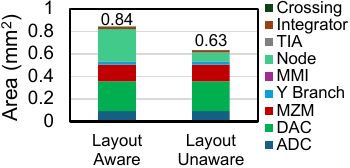}
    \label{fig:AreaBreak}}
    \subfloat[]{\includegraphics[width=0.395\columnwidth]{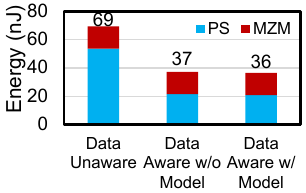}
    \label{fig:Data-Aware}}
    \vspace{-8pt}
    \caption{\small (a) TeMPO~\cite{NP_JAP2024_Zhang} area breakdown with layout awareness.
    (b) SCATTER~\cite{NP_ICCAD24_Gu} energy breakdown with data awareness.}
    \label{fig:PhysicalAware}
    \vspace{-10pt}
\end{figure}

%% file: fig_tex/fig_vgg8_mapping.tex
\begin{figure}
    \centering
    \vspace{-5pt}
    \includegraphics[width=0.75\columnwidth]{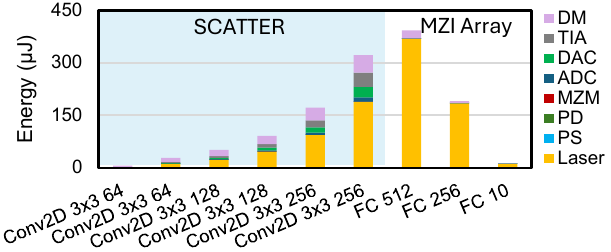}
    \vspace{-5pt}
    \caption{\small Layer energy breakdown with heterogeneous layer mapping of VGG-8(CIFAR10)~\cite{NN_COMPGRAPHSTAT2013_Simon}.
    Convolutions are mapped to SCATTER~\cite{NP_ICCAD24_Gu}, and Linear layers are mapped to MZI meshes~\cite{NP_NATURE2017_Shen}.}
    \label{fig:Vgg8}
    \vspace{-10pt}
\end{figure}

%% file: doc/5conclusion.tex
\vspace{-7pt}
\section{Conclusion}
\vspace{-3pt}
\label{sec:conclusion}
This paper presents \SIM, a cross-layer modeling and simulation framework for EPIC AI hardware, enabling photonic-specific design evaluation and fair comparisons across implementations.
By emphasizing accurate device/circuit-level modeling with generic and extensible representations, \SIM bridges hardware and software stacks to support flexible hardware construction, validation, and architecture exploration with multi-dimensional metric trade-offs.
We have validated \SIM's accuracy against prior work and will continue expanding its capabilities to help researchers uncover insights and drive innovation in high-performance, energy-efficient photonic computing systems.